\documentclass[twocolumn]{aastex63}
\newcommand{\new}[1]{#1}

\usepackage[scaled]{helvet}
 
\usepackage[T1]{fontenc}

\submitjournal{ApJL}

\shorttitle{Carbon fractionation from disk and planetesimal processing}
\shortauthors{Lichtenberg \& Krijt}

\begin{document}

\title{
System-level fractionation of carbon from disk and planetesimal processing
}

\correspondingauthor{Tim Lichtenberg}
\email{tim.lichtenberg@physics.ox.ac.uk}

\author[0000-0002-3286-7683]{Tim Lichtenberg}
\affiliation{Atmospheric, Oceanic and Planetary Physics, Department of Physics, University of Oxford, Oxford OX1 3PU, UK}

\author[0000-0002-3291-6887]{Sebastiaan Krijt}
\affiliation{School of Physics and Astronomy, University of Exeter, Stocker Road, Exeter EX4 4QL, UK}

\begin{abstract}
Finding and characterizing extrasolar Earth analogs will rely on interpretation of the planetary system's environmental context. The total budget and fractionation between C--H--O species sensitively affect the climatic and geodynamic state of terrestrial worlds, but their main delivery channels are poorly constrained. We connect numerical models of volatile chemistry and pebble coagulation in the circumstellar disk with the internal compositional evolution of planetesimals during the primary accretion phase. Our simulations demonstrate that disk chemistry and degassing from planetesimals operate on comparable timescales and can fractionate the relative abundances of major water and carbon carriers by orders of magnitude. As a result, individual planetary systems with significant planetesimal processing display increased correlation in the volatile budget of planetary building blocks relative to no internal heating. Planetesimal processing in a subset of systems increases the variance of volatile contents across planetary systems. Our simulations thus suggest that exoplanetary atmospheric compositions may provide constraints on \emph{when} a specific planet formed.
\end{abstract}


\section{Introduction} \label{sec:intro}

The surface and climatic conditions of rocky planets are sensitive to the volatile budget inherited from planetary formation. Variations in the main channels of volatile delivery and reprocessing during primary accretion affect the composition and thus redox state of planetary mantles, which in turn control atmospheric speciation, and hence surface geochemistry and long-term climate \citep{2021SSRv..217...22G,Graham21}. In particular young planets, comparable to the Hadean Earth and Pre-Noachian Mars, are sensitively affected by both the delivery during primary accretion, and potential secondary volatile delivery \new{from scattering and} dynamical transfer \new{of left-over planetesimals}, after the primary accretion phase has ceased \citep{Raymond2020PlanetaryAstrobiology}. Late volatile delivery can shift the atmospheric composition during transient epochs in the direction of the impactor composition: volatile-rich, metal-poor impactors oxidize Hadean-like atmospheres, volatile-poor, iron-rich impactors reduce the atmosphere; the latter is regarded a favorable environment for subaerial prebiotic chemistry \citep{Benner2019,2020PSJ.....1...11Z,2020AsBio..20.1476F}.

The heritage and processing of volatile elements during their journey from the interstellar medium (ISM) to the atmospheres of rocky planets, however, is poorly understood. Gas and grain-surface chemistry set the speciation and dominant carriers of hydrogen, carbon, oxygen, nitrogen, and sulfur before and during the infall of the gas and dust phase onto the disk, where coagulation, fragmentation, and settling processes redistribute ice phases in the outer parts of the disk \citep{oberg2011,obergbergin2021}. The depletion of mm-sized dust grains in ALMA disks suggests that the onset of the accretion phase operates fast, and a major fraction of the initial dust mass is depleted toward later disk stages \citep{2016ApJ...828...46A}. Isotopic ratios in Solar System planetary materials suggest that a fraction of volatiles in the terrestrial planets was inherited from thermally pristine sources \citep{2014Sci...345.1590C}, but recent evidence highlights the role of local \citep{Piani+2020,2021NatAs.tmp...14G} and originally volatile-rich, thermally reprocessed materials as possible carriers of water \citep{2019NatAs...3..307L,2021Sci...371..365L}, carbon species \citep{Hirschmann21}, and moderately volatile elements \citep{2019GeCoA.260..204S} during accretion. 

\new{Conversely, (small) comets and Kuiper Belt objects represent populations of icy bodies that formed in the outer Solar System, with volatile abundances closer to pre- and protostellar chemistry \citep{mumma_charnley2011, drozdovskaya2019, grundy2020}, and may have largely escaped significant thermal processing \citep{2019Icar..326...10B,Golabek21}.}

\section{Volatile processing in the disk vs. planetesimals} \label{sec:methods}

\begin{figure*}[tb]
\centering
\includegraphics[width=.90\textwidth]{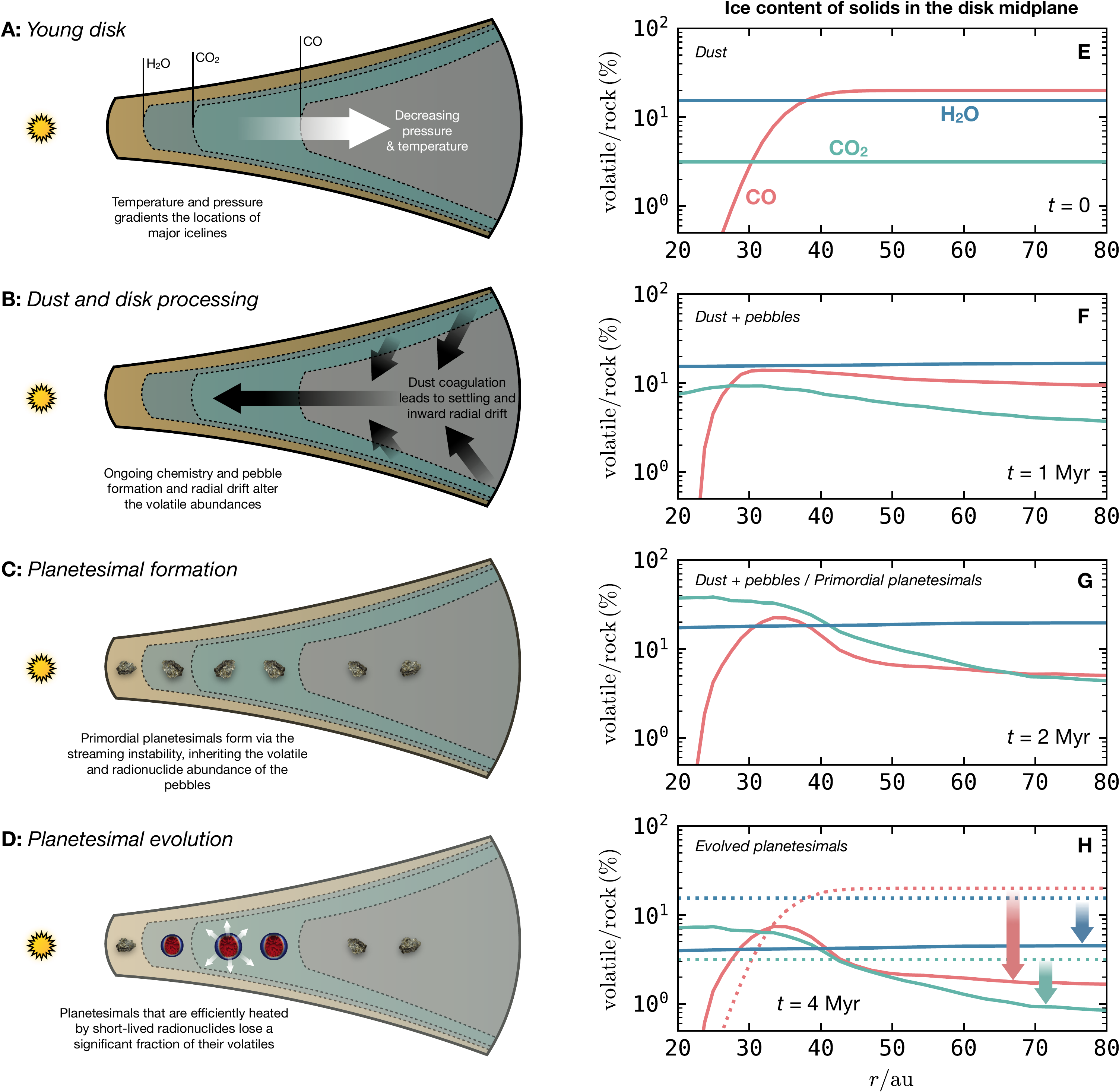}
\caption{\textsf{\emph{Left:} Sketch illustrating the sequence of events during which we follow the fractionation of H$_2$O, CO, and CO$_2$. \emph{Right:} \new{Volatile/rock mass ratios for:} dust near the midplane (E), dust + pebbles (F), and primordial (G) and evolved (H) planetesimals as a function of radius in our fiducial model run. In (H), the dotted lines depict the initial conditions at $t=0$.}}
\label{fig:fig2}
\end{figure*}

\begin{figure*}[tb]
\centering
\includegraphics[width=.49\textwidth]{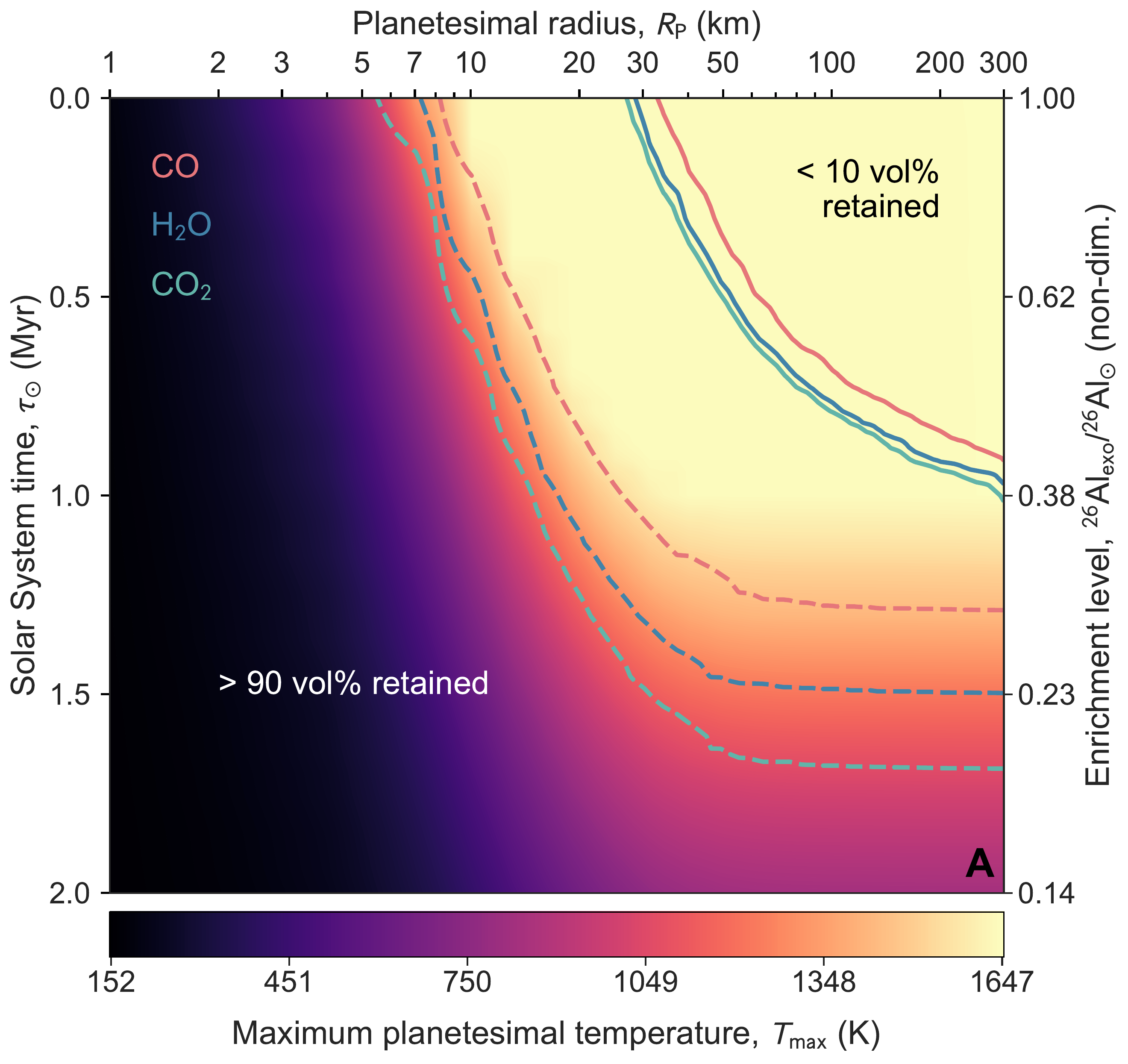}
\includegraphics[width=.49\textwidth]{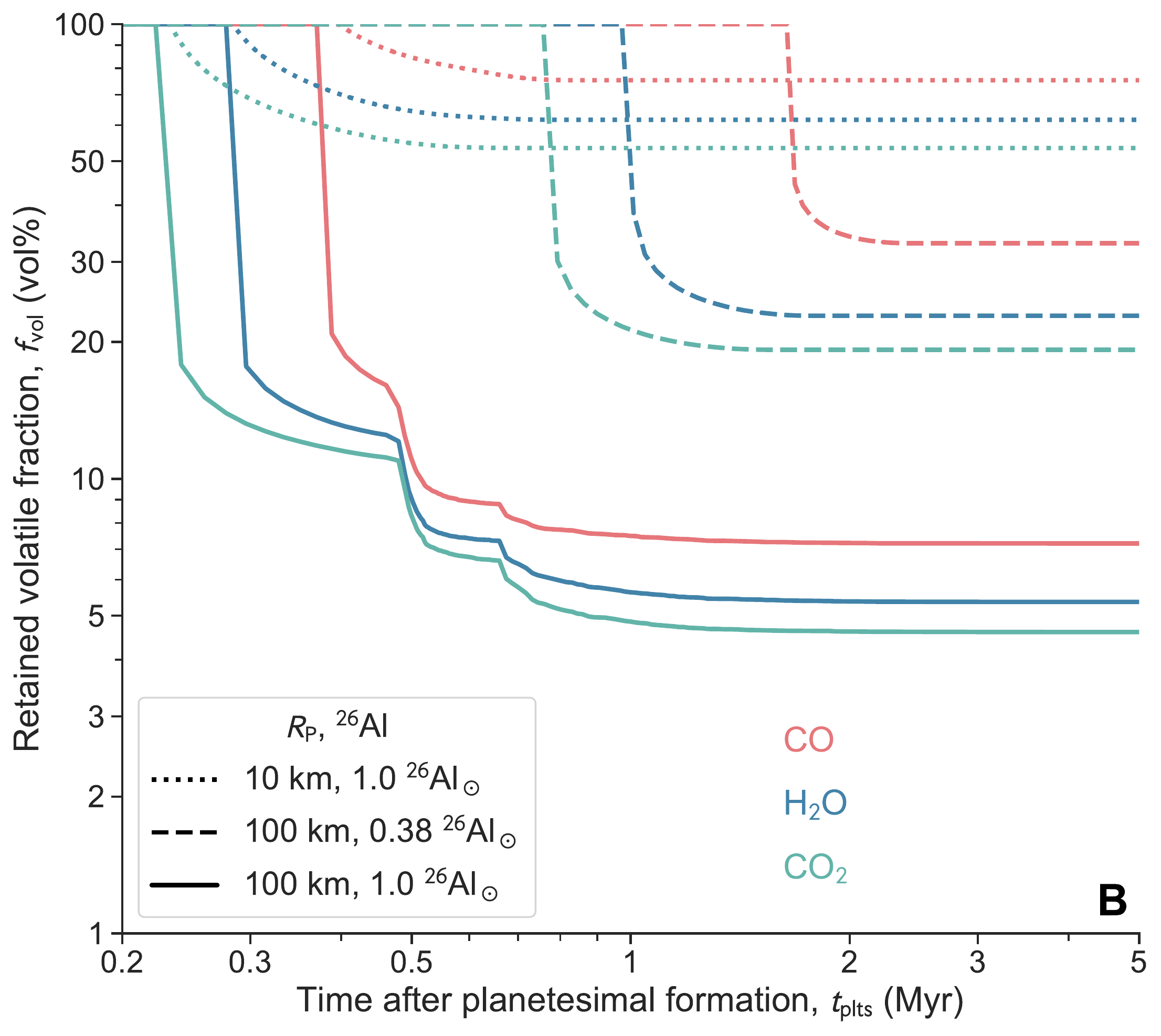}
\caption{\textsf{Thermal evolution of planetesimals and loss of H$_2$O, CO$_2$, and CO. \emph{(Left)} Peak temperatures in the interior of planetesimals from radioactive decay of $^{26}$Al (background color), interpolated from a grid of planetesimal simulations. The y-axes show enrichment level at the time of planetesimal formation ($^{26}$Al$_\mathrm{exo}/^{26}$Al$_{\Sol}$) and time after CAI formation in the Solar System. The x-axis (top) gives the radius of planetesimals. Colored lines indicate the retention of $>$ 90 vol\% (dashed) or 10 vol\% (solid) of H$_2$O (blue), CO$_2$ (green), and CO (red) after peak heating inside the planetesimal. \emph{(Right)} Loss  of H$_2$O (blue), CO$_2$ (green), and CO (red) over time for different combinations of $^{26}$Al abundance and planetesimal radius.}}
\label{fig:fig1}
\end{figure*}

In this Letter, we follow the abundances of H$_2$O, CO$_2$, and CO starting from microscopic icy grains in the outer regions of protoplanetary disks, through pebbles, to primordial and ultimately evolved/processed planetesimals. Here we briefly describe the relevant processes before and after planetesimal formation, which are illustrated in Fig.~\ref{fig:fig2}A--D. Our focus is on the disk region straddling the CO iceline.

\subsection{Disk processing: dynamics and chemistry in gas and solid phase}\label{sec:disk_processing}
The volatile content of the outer parts of protoplanetary disks is largely inherited from the cold molecular cloud. Typically, the majority of the volatile carbon and oxygen is carried by H$_2$O, CO$_2$, and CO \citep[e.g.,][]{obergbergin2021}. For such a composition, the volatile content of the microscopic dust grains is then set by their location relative to major icelines, imaginary boundaries that separate regions where specific molecules are stable as ices \citep[Fig.~\ref{fig:fig2}A and][]{oberg2011}. However, chemical processing and the formation and distinct dynamical behavior of pebble-size particles is expected to alter the abundances of key volatile species both locally and on disk-wide scales \citep[][and references therein]{bosman2018,krijt2018,krijt2020,obergbergin2021}. Here, we use the results of \citet{krijt2020}, who presented numerical models merging chemical processing of CO, dust coagulation, and pebble dynamics, to describe this stage. While these models were developed with the goal of studying the inferred depletion of gas-phase CO in the upper disk regions \citep[][]{zhang2019}, they offer a unique window into the evolution of the volatile content of solids growing and moving in the disk midplane. Figure \ref{fig:fig2} shows the radial variation in the volatile content (plotted as \new{mass} content relative to the refractory dust component, which can in some cases exceed 100\%) of: dust at $t=0$ (panel E); dust + pebbles after 1~Myr (panel F) and 2~Myr (panel G). \new{For comparison, water is the dominant volatile seen in comets, with CO and CO$_2$ abundances varying between ${\approx}1-30\%$ that of H$_2$O \citep{mumma_charnley2011}. Variations in volatile composition (of these and other species) may be linked to dynamical type and formation time or location \citep[e.g.,][]{dellorusso2016,eistrup2019}.} \new{Panels F and G of Fig.~\ref{fig:fig2} highlight several effects of disk processing:} \emph{(i)} a decrease in CO and CO$_2$ in the outer disk regions because of chemical processing and transfer to CH$_4$ and CH$_3$OH ice \citep[][]{bosman2018,krijt2020}, and \emph{(ii)} an enhancement of CO around the CO iceline, resulting from the retro-diffusion of CO vapor brought in through radial drift \citep[][]{stammler2017,krijt2018}. In this formulation, the timescales for the chemical processing are determined mainly by the cosmic ray ionisation rate \citep{bosman2018}, set here to $\zeta_\mathrm{CR}=10^{-17}\mathrm{~s^{-1}}$.

\subsection{Planetesimal formation}
In recent years the streaming instability has emerged as a leading contender to overcome the various barriers to growth from dust grains to the first planetesimals in protoplanetary disks \citep[][]{2014prpl.conf..547J}. 
Evidence from the asteroid and Kuiper belt \new{\citep{2017Sci...357.1026D,2019Sci...363..955S,mckinnon2020} and fluid dynamical simulations \citep{2019ApJ...885...69L}} suggest the planetesimal population to be dominated by $\sim$\new{10--100} km bodies. The composition of these bodies will closely follow the local composition of solids at the time of gravitational collapse, but the timing and location of planetesimal formation may vary considerably \citep{2017AA...602A..21S,2019ApJ...884L...5S}. The model of \citet{krijt2020} does not explicitly model the formation of solids beyond pebble sizes, and for the purpose of this work we treat the formation time $\tau_\mathrm{pf}$ and radii $R_\mathrm{p}$ of planetesimals to be free model parameters.

\subsection{Planetesimal processing: melting and degassing}\label{sec:plts_processing}
We simulate the devolatilization and degassing from planetesimals that are internally heated by the decay of the short-lived radionuclide $^{26}$Al as reviewed in \citet{Monteux2018}. We use the thermal evolution simulations from \citet{2021Sci...371..365L}, with degassing thresholds for the volatile release of H$_2$O, CO$_2$, and CO from the planetesimal interior. We \new{parameterize} planetesimal devolatilization based on laboratory experiments on volatile release from heated chondrite samples. We assume complete release of CO$_2$ at 800$^{\circ}$C (1073 K) \citep{1995Metic..30..639M} and of H$_2$O at 950$^{\circ}$C (1223 K) \citep{1999CoMP..135...18N}. Minor abundances of CO may be regenerated by smelting between graphite and ferrous silicates until the mobilization of the first silicate melts \citep{2014EPSL.390..128F} at a silicate melt fraction of $\varphi \approx$ 0.1 \citep{2019EPSL.507..154L} (${\approx}1470$ K at 1 bar), which we define as the threshold for CO release. \new{The above thresholds define upper limits on devolatilization of hydrated and carbonated rocks that can be (re-)generated by fluid-rock interactions in the host body during heating. Ice melting and degassing commence earlier and at lower temperatures, but we focus here on conservative upper limits when even left-over materials are essentially devoid of any volatile form of carbon.} 

We focus on planetesimals with $\approx$ 100 km in radius, but also explore the evolution of planetesimals with 10 km radius. The distribution of initial $^{26}$Al abundances between planetary systems is expected to vary by orders of magnitude, depending on the birth star-forming environment of a given planetary system \citep{2020RSOS....701271P,2020A&A...644L...1R}. Compositional changes due to $^{26}$Al heating at the time of planetesimal formation are significant within about 3 half-lives of the initial Solar System value ($^{26}$Al$_{\Sol}$) \citep{2019NatAs...3..307L}, which is why we limit our simulations here to $^{26}$Al abundances at the time of planetesimal formation in the range of $^{26}$Al$_\mathrm{exo}$ $\in$ [0.1, 1.0] $^{26}$Al$_{\Sol}$.

Higher $^{26}$Al and larger body size lead to higher temperatures and increasing loss of H$_2$O, CO$_2$, and CO to space (Fig. \ref{fig:fig1}A). Variability for bodies larger than $\sim$100 km is dominated by changes in $^{26}$Al levels and hence system enrichment at planetesimal formation. Volatile retention in Fig. \ref{fig:fig1}A is set by the devolatilization thresholds: loss efficiency decreases from CO$_2$ to H$_2$O to CO. In the high temperature regime, differences arise from retention in the thin conductive crusts of planetesimals. In the lower parts and at lower heating rates, the different thresholds separate the devolatilization trends for the different species more visibly: 10\% loss rates (90\% retention) are offset between 0.15 to 0.3 $^{26}$Al$_{\Sol}$. Volatile loss is variable and sensitive to the planetesimal formation conditions \citep[Fig. \ref{fig:fig1}B, cf.][]{2019NatAs...3..307L,2021Sci...371..365L}. We here compare loss rates for 10 and 100 km planetesimal radii and enrichment levels of 1.0 and 0.38 $^{26}$Al$_{\Sol}$. At 1.0 $^{26}$Al$_{\Sol}$, 10 km and 100 km radii planetesimals start losing their volatiles at roughly the same time, about 0.2 to 0.4 Myr after planetesimal formation, but larger planetesimals dehydrate more efficiently. 10 km planetesimals retain pristine materials in 60 to 80 vol\% of their interiors, 100 km planetesimals retain on the order of 5 to 8 vol\%. At 0.38 $^{26}$Al$_{\Sol}$, 100 km planetesimal (dashed lines) degas later, between 0.8 to 1.5 Myr after planetesimal formation, and loss efficiency decreases. Between 20 to 40 vol\% of pristine materials can be retained.

\section{Combined trends} \label{sec:results}

\begin{figure*}[tb]
\centering
\includegraphics[width=.95\textwidth]{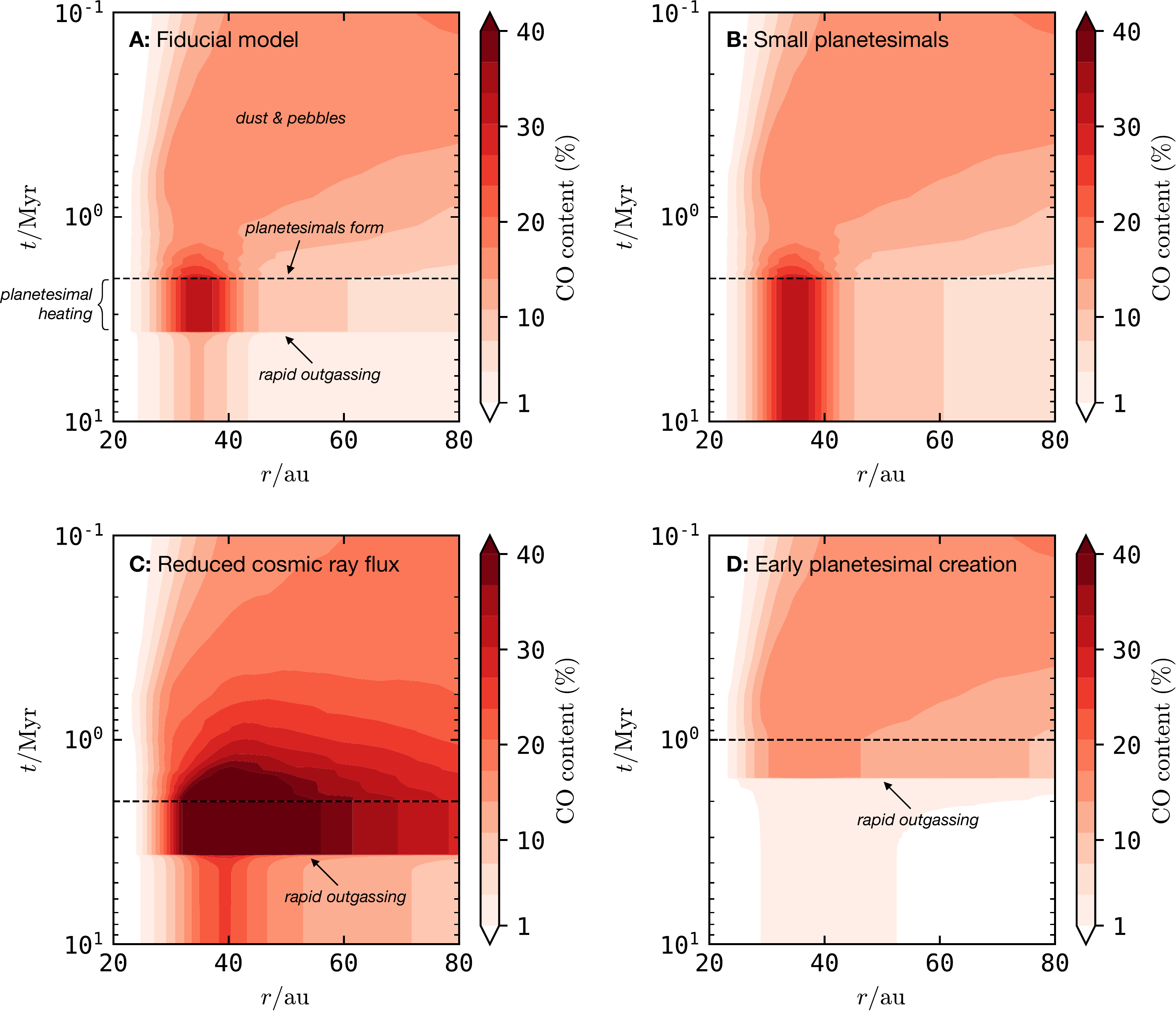}
\caption{\textsf{CO content \new{(defined as in Fig.~\ref{fig:fig2})} of dust and pebbles in the midplane (above the dashed line) and planetesimals (below the dashed line) as a function of heliocentric distance and time for 4 different models. (A): Fiducial model, with $R_\mathrm{p}=100\mathrm{~km}$, $\tau_\mathrm{pf}=2\mathrm{~Myr}$, $\zeta_\mathrm{CR}=10^{-17}\mathrm{~s^{-1}}$, and $^{26}$Al$_\mathrm{exo}/^{26}$Al$_{\Sol}=0.38$. (B): Smaller planetesimals with $R_\mathrm{p}=10\mathrm{~km}$. (C):  $\zeta_\mathrm{CR}=10^{-18}\mathrm{~s^{-1}}$. (D): Earlier planetesimal formation and a higher radionuclide content with $\tau_\mathrm{pf}=1\mathrm{~Myr}$ and $^{26}$Al$_\mathrm{exo}/^{26}$Al$_{\Sol}=0.75$.}}
\label{fig:fig3}
\end{figure*}

We investigate how degassing from planetesimals (Sect.~\ref{sec:plts_processing}) affects the volatile evolution \new{inherited} from disk-based solid- and gas-phase chemistry and dust coagulation (Sect.~\ref{sec:disk_processing}, Fig.~\ref{fig:fig2}E-F). In our fiducial model we assume 100 km planetesimals form after 2~Myr, with $^{26}$Al$_\mathrm{exo}/^{26}$Al$_{\Sol}=0.38$ at the time of their formation. By feeding the midplane compositions from the disk model (at $t=2~\mathrm{Myr}$) to the planetesimal evolution calculation, we compute the evolution of the H$_2$O, CO$_2$, and CO content of the planetesimals. The results of the fiducial model are shown in Fig.~\ref{fig:fig2}H. For these parameters a significant fraction of each volatile is lost during outgassing. The final water content is fairly constant with disk location (around 4\%), while the CO$_2$ content shows a decline with orbital distance. The CO content peaks at 10\% between 30--40 au, preserving the maximum in CO ice content that was formed as a result of pebble drift. In contrast, planetesimals formed around 20 au (i.e., inside the CO iceline) contain no CO, but close to 10\% CO$_2$. Thus, the interplay of the various processes described here (chemical processing, pebble drift, and planetesimal outgassing) can substantially alter not just the individual but also the \emph{relative} abundances of major carbon and oxygen carriers.

The complete time evolution of the fiducial model is shown in Fig.~\ref{fig:fig3}A. Here, focusing exclusively on CO, we show \new{that} the volatile abundance varies in both space and time. The region above the dashed lines depicts dust and planetesimals, and the region below it exclusively planetesimals. We can clearly distinguish 3 distinct periods in time. First, chemical processing and dust evolution alter the volatile content ($t<2\mathrm{~Myr}$). Then, planetesimals form, effectively quenching the composition of solids at the time of their formation ($t\approx 2-3.5\mathrm{~Myr}$). As planetesimals heat up, however, they rapidly lose a significant fraction of their CO content (Fig.~\ref{fig:fig1}B), leaving them relatively CO poor ($t > 3.5\mathrm{~Myr}$).

Figure \ref{fig:fig3}B--D presents a parameter exploration showing the effects of varying three key parameters in our model. First, assuming planetesimals are formed smaller (10~km) enables them to avoid efficient heating, resulting in virtually no CO loss after planetesimal formation (panel B). Lowering the cosmic ray ionisation rate by a factor of 10 relative to the fiducial case reduces the amount of CO that is processed on grain surfaces, resulting in a even higher CO content just outside the CO iceline (panel C). However, if planetesimals form earlier (1~Myr) and with a higher radionuclide content ($^{26}$Al$_\mathrm{exo}/^{26}$Al$_{\Sol}=0.75$), there is less time for the CO peak to develop during the dust evolution phase \emph{and} planetesimals degas more readily, resulting in very low final CO contents (panel D).

\section{Volatile fractionation of planetary systems} \label{sec:discussion}

\begin{figure*}[tb]
\centering
\includegraphics[width=.99\textwidth]{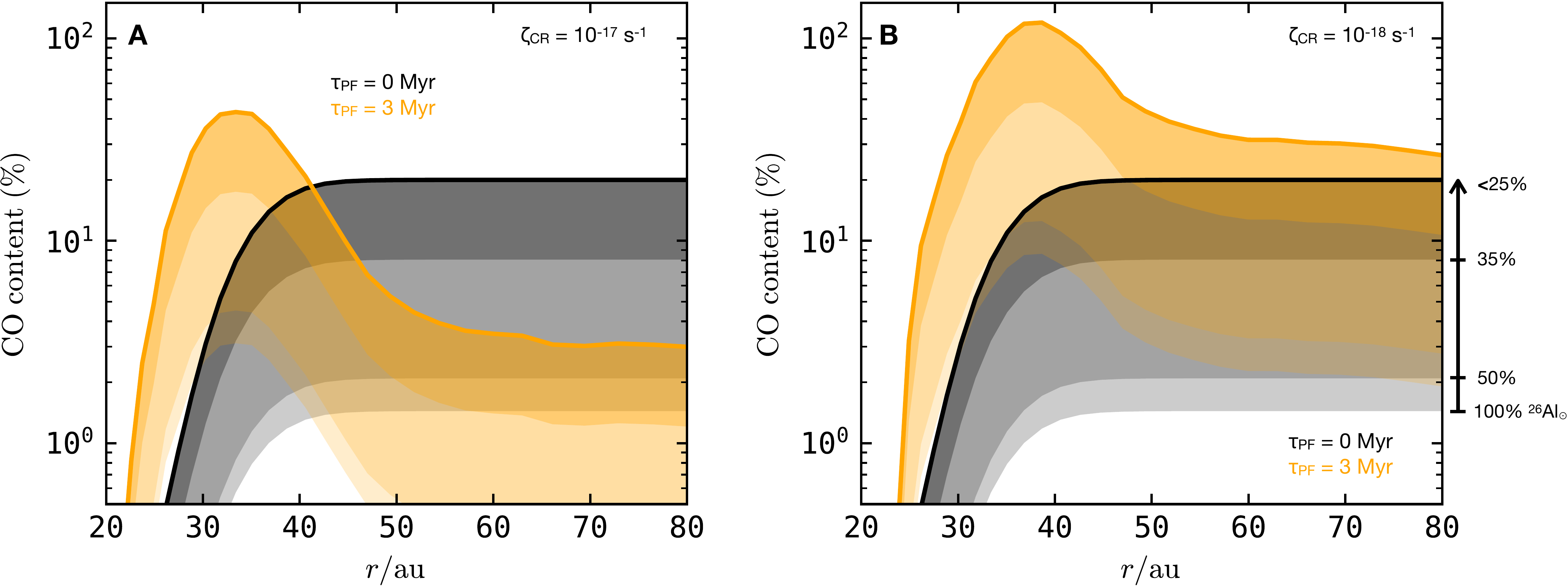}
\caption{\textsf{Profile of \new{CO content} of planetesimals \emph{after} degassing as a function of heliocentric distance for models without any chemical processing or pebble drift during the disk phase (black) compared to those with 3 Myr or ongoing chemistry and dust evolution (orange). Different shadings depict different levels of $(^{26}$Al$_\mathrm{exo}/^{26}$Al$_{\Sol})$ at the time of planetesimal formation, and panels A and B differ in the cosmic ray ionisation rate.}}
\label{fig:fig4}
\end{figure*}

In Fig.~\ref{fig:fig4} we plot the radial profile of the final planetesimal CO abundance for planetesimals that form effectively at $t=0$ (i.e., before any chemical processing or dust/pebble transport can occur in the gas-rich disk) and after $t=3~\mathrm{Myr}$ (when significant evolution has taken place). The shading indicates different levels of $(^{26}$Al$_\mathrm{exo}/^{26}$Al$_{\Sol})$ at the time of planetesimal formation, and we compare two disk models with different cosmic ray abundances. All models are for $R_\mathrm{p}=100~\mathrm{km}$. Several things stand out, in particular: \emph{(i)} in disks that experienced significant radial drift across the CO iceline, the radial profile of planetesimal CO content shows a prominent peak just exterior to the location of the CO iceline. The width of this peak here is about 10--20 au, but can vary depending on the strength of radial diffusion \citep[][]{stammler2017,krijt2018}. \new{This effect can potentially explain the existence of planetesimals or comets with anomalously high CO/H$_2$O ratios \citep{price2021,mousis2021}. More typical comets also display significant variation in CO/H$_2$O ratios in their comae, although it is not clear whether this represents variation in formation conditions or later dynamical evolution and processing \citep{dellorusso2016, altwegg2019}}.   \emph{(ii)} In disks with a higher cosmic ray rate, the CO content of planetesimals will be lower, especially if these planetesimals form after a few million years. \emph{(iii)} The $^{26}$Al content plays a major role in shaping the final abundance and fractionation of volatiles, resulting in order-of-magnitude variations for the ranges explored here. 

In a broader context, our results demonstrate the necessity of considering volatile fractionation both before and immediately after planetesimal accretion when attempting to connect the volatile budget of protoplanetary disk midplanes to those of planetary-scale objects. These results illustrate that from a planetary systems perspective, the anticipated fractionation effects between individual volatiles--here specifically illustrated for the major carbon-bearing species--from disk plus planetesimal processing differ significantly from those expected from pure disk-based processing alone \citep[e.g.][]{oberg2011}. We here discuss this result in the context of planet formation theory, and relate it to potentially observable signatures in extrasolar planetary systems that connect the volatile budget of exoplanets to the specific accretion pathway of the Solar System.

\subsection{Timing and mode of volatile accretion}

Planetary growth dominated by mutual accretionary collisions between planetesimals extends well beyond the lifetime of the gaseous protoplanetary disk \citep{Raymond2020PlanetaryAstrobiology}. The direct accretion of pebbles from the ambient disk onto growing protoplanets may shift the phase of primary mass addition to earlier stages during the disk stage itself \citep{2017ASSL..445..197O}. Protoplanets accreted within the disk lifetime undergo orbital migration and inherit a substantial amount of their mass from beyond the snowline or other icelines \citep{Venturini20}. Our simulations in this work demonstrate that the major volatile carbon carriers CO and CO$_2$ are affected \new{by outgassing from planetary precursors} to a similar degree as water \citep{2019NatAs...3..307L}. Planetary growth beyond the disk stage inherits the volatile budget from such devolatilized bodies \citep{2015ApJ...804....9C} because the degassing from planetesimals operates on much shorter timescales than collisional growth after the disk phase. CO and CO$_2$ are a major ingredient for the mantle composition, surface conditions, and climate of terrestrial worlds inherited from planet formation, but the Earth is carbon-poor relative to the nominal values in the ISM \citep{2015PNAS..112.8965B} and the earliest known planetesimal bodies in the Solar System display carbon fractionation trends indicative of extensive devolatilization \citep{Hirschmann21} and carbon-rich fluid mobilization \citep{Tsuchi21}. Planetesimal degassing as illustrated in the presented simulations limits the total amount of carbon that can be delivered to nascent planets by decreasing the volatile abundances of primitive, ice-rich planetesimals to system-wide comparable levels and flattens the volatile variability in planetary building blocks (Fig.~\ref{fig:fig4}) that can contribute to the primordial atmospheric composition of protoplanets. \new{This suggests that left-over material in substantially processed planetesimal generations converges toward the abundances of refractory forms of carbon inherited from the disk \citep{Li+21}.} 

In addition to volatile delivery during primary accretion, \new{planetesimal impacts during the tail-end of accretion} have been suggested as a source of volatiles for the early Earth. The bulk compositions of Earth \citep{hirschmann2016constraints} and Venus \citep{2020NatGe..13..265G}, however, suggest the nature of the late accreted material to be dry in composition and chemically reduced. Our results suggest that the level of degassing from planetesimals acts to limit the total mass of volatiles accreted from bodies that are scattered from the outside to inner orbits of planetary systems. This alters the prospects for creating transiently reducing conditions on Hadean Earth analogs \citep{Benner2019,2020PSJ.....1...11Z}, and suggests that the reducing power and carbon budget of late \new{accretion} episodes \citep{2017ApJ...843..120S} is directly tied to the geophysical and geochemical evolution of the planetesimals from the \new{combined} effects of disk chemistry and internal heating.

\subsection{Debris disks \& Kuiper belt analogs}

Another connection exists between the planetesimals formed in our simulations and left-over icy bodies in the Solar System and beyond \citep[][Sect.~4.6]{krijt2020}. Evidence suggests bodies in the Kuiper Belt formed late \citep{2019Icar..326...10B}, with little $^{26}$Al and hence minor heating \citep{Golabek21} -- similar to the end-member case in Fig.~\ref{fig:fig3}B. Given the expected variation in $^{26}$Al content from system to system, it is interesting to look at extra-solar Kuiper Belts \citep{wyatt2020}, as it is now becoming possible to constrain the volatile content of planetesimals at 10s to 100s of au in a handful of systems \citep[e.g.,][]{marino2016,kral2017}. Constraints on the (CO+CO$_2$) mass content of planetesimals range from ${<}1$ to ${\approx}35\%$ \citep{wyatt2020}, broadly consistent with the ranges we find in e.g., Figs.~\ref{fig:fig2} and \ref{fig:fig3}. If the debris disk ring location, volatile content, and planetesimal size \citep[e.g.,][]{krivov2021} can be sufficiently constrained, models such as the ones presented here can be used to infer planetesimal formation timescales and the extent of past disk processing for these systems.

\new{Finally, interstellar interlopers offer a unique window in to the volatile content of extra-solar planetesimals. In the framework presented here, the high CO abundance of 2I/Borisov \citep{cordiner2020} points to formation (just) outside the CO iceline, and only minimal chemical processing and outgassing after formation (e.g., Fig.~\ref{fig:fig4}).}

\subsection{Environmental information from short-period exoplanets}

Exoplanet surveys that will seek to detect and characterize rocky extrasolar planets \citep{2019AJ....158...83A,2021arXiv210107500L,2021arXiv210308481H} will rely on environmental clues to inform and constrain the composition and history of planetary systems. Accounted for secondary loss processes, the composition of short-period exoplanets can yield crucial information on the volatile content of the planetary system as a whole. For instance, assuming radial variation in volatile content alone, the absence of large volatile quantities in super-Earths has been suggested to result from dry formation inside the snowline \citep{2020PSJ.....1...36K}. However, as we show in this letter, formation under high abundances of short-lived radioactive isotopes can devolatilize the planetary building blocks, and hence reconcile orbital migration with volatile-poor compositions. Enhanced distribution statistics on the bulk composition of small exoplanets below the Kepler radius valley will further provide information on the diversity of formation paths, and hence volatile sources, of planets on a population level \citep{2019PNAS..116.9723Z,2021MNRAS.tmp..530R}. In addition, prolonged magma ocean phases due to runaway greenhouse climates on short periods  \citep{2015ApJ...806..216H} and on wider orbits around young stars \citep{2019AA...621A.125B} promise to reveal characteristic signatures depending on their volatile content \citep{Lichtenberg2021JGRP}.

While hot rocky exoplanets provide means to probe the detailed inventory of volatiles, sub-Neptunes and Hot Jupiters are the more tangible targets in the upcoming decade. Their atmospheric metallicity informs the distribution across systems that are dynamically altered by planetary migration and hence probe the volatile content of planetary building blocks beyond the snowline \citep{2019MNRAS.482.1485P}. Based on radial and disk chemistry variation of volatile content alone, these systems should show a wide spread in atmospheric speciation \citep{booth2017,2020MNRAS.499.2229N}. If internal processing on planetesimals, like suggested here, quenches the initially accreted volatile content, then volatile trends in gaseous exoplanets should follow system-wide trends, altering the distribution statistics from system to system.

\section{Summary} \label{sec:conclusions}

In this work, we simulated the time- and location-dependent evolution of H$_2$O, CO, and CO$_2$ from combined disk and planetesimal processing. At a single location the volatile bulk compositions of planetesimals vary by orders of magnitude over time from disk chemistry alone. Heating and degassing by planetesimals, however, lead to rapid loss of volatiles on a timescale comparable to disk-based chemical evolution and planetary accretion. Solar System-like enrichment levels of the short-lived radioactive isotope $^{26}$Al lead to relative fractionation between individual volatiles by up to two orders of magnitude, depleting planetary volatile carriers in H$_2$O, CO, and CO$_2$ down to levels of ${<}1\%$. Importantly, the relative fractions of H$_2$O, CO, and CO$_2$ can be inverted on time intervals relevant to planetary accretion.

The degree of volatile fractionation suggested in this work alters the prospects for the climatic and surface conditions of rocky planets. If planetesimal formation is rapid, carbon-bearing species fractionate on a system-wide level as a function of the internal heating in planetesimals. These effects alter the potential for early volatile delivery and during late \new{accretion} episodes, and hence shape total carbon budget, redox state, and primitive surface conditions of young rocky planets during the first few hundred million years. Our results can be tested via population statistics of short-period exoplanets, atmospheric speciation of molten exoplanets, and debris disks, and provide contextual information on the viability of Earth-like climates on temperate orbits within individual and across planetary systems.

\vspace{0.15cm}
\small{
\emph{Acknowledgments:} % \acknowledgments
We thank Gregor Golabek and an anonymous reviewer for constructive comments on the manuscript. T.L. was supported by a grant from the Simons Foundation (SCOL award No. 611576). This work benefitted from information exchange within the program `Alien Earths' (NASA Grant No. 80NSSC21K0593) for NASA's Nexus for Exoplanet System Science (NExSS) research coordination network.

\vspace{0.15cm}
\emph{Software:} \textsc{i2elvis} \citep{2007PEPI..163...83G}, \textsc{numpy} \citep{Numpy2020}, \textsc{scipy} \citep{2020SciPy-NMeth}, \textsc{pandas} \citep{pandas:2010}, \textsc{matplotlib} \citep{Hunter:2007}, \textsc{seaborn} \citep{seaborn:2018}.
}

%\appendix
%\section{}

\bibliography{references}{}

\begin{thebibliography}{}
\expandafter\ifx\csname natexlab\endcsname\relax\def\natexlab#1{#1}\fi
\providecommand{\url}[1]{\href{#1}{#1}}
\providecommand{\dodoi}[1]{doi:~\href{http://doi.org/#1}{\nolinkurl{#1}}}
\providecommand{\doeprint}[1]{\href{http://ascl.net/#1}{\nolinkurl{http://ascl.net/#1}}}
\providecommand{\doarXiv}[1]{\href{https://arxiv.org/abs/#1}{\nolinkurl{https://arxiv.org/abs/#1}}}

\bibitem[{{Altwegg} {et~al.}(2019){Altwegg}, {Balsiger}, \&
  {Fuselier}}]{altwegg2019}
{Altwegg}, K., {Balsiger}, H., \& {Fuselier}, S.~A. 2019, \araa, 57, 113,
  \dodoi{10.1146/annurev-astro-091918-104409}

\bibitem[{{Ansdell} {et~al.}(2016){Ansdell}, {Williams}, {van der Marel},
  {Carpenter}, {Guidi}, {Hogerheijde}, {Mathews}, {Manara}, {Miotello},
  {Natta}, {Oliveira}, {Tazzari}, {Testi}, {van Dishoeck}, \& {van
  Terwisga}}]{2016ApJ...828...46A}
{Ansdell}, M., {Williams}, J.~P., {van der Marel}, N., {et~al.} 2016, \apj,
  828, 46, \dodoi{10.3847/0004-637X/828/1/46}

\bibitem[{{Apai} {et~al.}(2019){Apai}, {Milster}, {Kim}, {Bixel}, {Schneider},
  {Liang}, \& {Arenberg}}]{2019AJ....158...83A}
{Apai}, D., {Milster}, T.~D., {Kim}, D.~W., {et~al.} 2019, \aj, 158, 83,
  \dodoi{10.3847/1538-3881/ab2631}

\bibitem[{{Benner} {et~al.}(2019){Benner}, {Bell}, {Biondi}, {Brasser},
  {Carell}, {Kim}, {Mojzsis}, {Omran}, {Pasek}, \& {Trail}}]{Benner2019}
{Benner}, S.~A., {Bell}, E.~A., {Biondi}, E., {et~al.} 2019, ChemSystemsChem,
  2, e1900035, \dodoi{10.1002/syst.201900035}

\bibitem[{{Bergin} {et~al.}(2015){Bergin}, {Blake}, {Ciesla}, {Hirschmann}, \&
  {Li}}]{2015PNAS..112.8965B}
{Bergin}, E.~A., {Blake}, G.~A., {Ciesla}, F., {Hirschmann}, M.~M., \& {Li}, J.
  2015, \pnas, 112, 8965, \dodoi{10.1073/pnas.1500954112}

\bibitem[{{Bierson} \& {Nimmo}(2019)}]{2019Icar..326...10B}
{Bierson}, C.~J., \& {Nimmo}, F. 2019, \icarus, 326, 10,
  \dodoi{10.1016/j.icarus.2019.01.027}

\bibitem[{{Bonati} {et~al.}(2019){Bonati}, {Lichtenberg}, {Bower}, {Timpe}, \&
  {Quanz}}]{2019AA...621A.125B}
{Bonati}, I., {Lichtenberg}, T., {Bower}, D.~J., {Timpe}, M.~L., \& {Quanz},
  S.~P. 2019, \aap, 621, A125, \dodoi{10.1051/0004-6361/201833158}

\bibitem[{{Booth} {et~al.}(2017){Booth}, {Clarke}, {Madhusudhan}, \&
  {Ilee}}]{booth2017}
{Booth}, R.~A., {Clarke}, C.~J., {Madhusudhan}, N., \& {Ilee}, J.~D. 2017,
  \mnras, 469, 3994, \dodoi{10.1093/mnras/stx1103}

\bibitem[{{Bosman} {et~al.}(2018){Bosman}, {Walsh}, \& {van
  Dishoeck}}]{bosman2018}
{Bosman}, A.~D., {Walsh}, C., \& {van Dishoeck}, E.~F. 2018, \aap, 618, A182,
  \dodoi{10.1051/0004-6361/201833497}

\bibitem[{{Ciesla} {et~al.}(2015){Ciesla}, {Mulders}, {Pascucci}, \&
  {Apai}}]{2015ApJ...804....9C}
{Ciesla}, F.~J., {Mulders}, G.~D., {Pascucci}, I., \& {Apai}, D. 2015, \apj,
  804, 9, \dodoi{10.1088/0004-637X/804/1/9}

\bibitem[{{Cleeves} {et~al.}(2014){Cleeves}, {Bergin}, {Alexand er}, {Du},
  {Graninger}, {{\"O}berg}, \& {Harries}}]{2014Sci...345.1590C}
{Cleeves}, L.~I., {Bergin}, E.~A., {Alexand er}, C. M.~O.~D., {et~al.} 2014,
  Science, 345, 1590, \dodoi{10.1126/science.1258055}

\bibitem[{{Cordiner} {et~al.}(2020){Cordiner}, {Milam}, {Biver},
  {Bockel{\'e}e-Morvan}, {Roth}, {Bergin}, {Jehin}, {Remijan}, {Charnley},
  {Mumma}, {Boissier}, {Crovisier}, {Paganini}, {Kuan}, \&
  {Lis}}]{cordiner2020}
{Cordiner}, M.~A., {Milam}, S.~N., {Biver}, N., {et~al.} 2020, Nat. Astron., 4,
  861, \dodoi{10.1038/s41550-020-1087-2}

\bibitem[{{Delbo} {et~al.}(2017){Delbo}, {Walsh}, {Bolin}, {Avdellidou}, \&
  {Morbidelli}}]{2017Sci...357.1026D}
{Delbo}, M., {Walsh}, K., {Bolin}, B., {Avdellidou}, C., \& {Morbidelli}, A.
  2017, Science, 357, 1026, \dodoi{10.1126/science.aam6036}

\bibitem[{{Dello Russo} {et~al.}(2016){Dello Russo}, {Kawakita}, {Vervack}, \&
  {Weaver}}]{dellorusso2016}
{Dello Russo}, N., {Kawakita}, H., {Vervack}, R.~J., \& {Weaver}, H.~A. 2016,
  \icarus, 278, 301, \dodoi{10.1016/j.icarus.2016.05.039}

\bibitem[{{Drozdovskaya} {et~al.}(2019){Drozdovskaya}, {van Dishoeck}, {Rubin},
  {J{\o}rgensen}, \& {Altwegg}}]{drozdovskaya2019}
{Drozdovskaya}, M.~N., {van Dishoeck}, E.~F., {Rubin}, M., {J{\o}rgensen},
  J.~K., \& {Altwegg}, K. 2019, \mnras, 490, 50, \dodoi{10.1093/mnras/stz2430}

\bibitem[{{Eistrup} {et~al.}(2019){Eistrup}, {Walsh}, \& {van
  Dishoeck}}]{eistrup2019}
{Eistrup}, C., {Walsh}, C., \& {van Dishoeck}, E.~F. 2019, \aap, 629, A84,
  \dodoi{10.1051/0004-6361/201935812}

\bibitem[{{Ferus} {et~al.}(2020){Ferus}, {Rimmer}, {Cassone},
  {Kn\&{\'\i}{\v{z}}ek}, {Civi{\v{s}}}, {{\v{S}}poner}, {Ivanek},
  {{\v{S}}poner}, {Saeidfirozeh}, {Kubel{\'\i}k}, {Dud{\v{z}}{\'a}k}, {Petera},
  {Juha}, {Pastorek}, {K{\v{r}}ivkov{\'a}}, \&
  {Kr{\^u}s}}]{2020AsBio..20.1476F}
{Ferus}, M., {Rimmer}, P., {Cassone}, G., {et~al.} 2020, Astrobiology, 20,
  1476, \dodoi{10.1089/ast.2020.2231}

\bibitem[{{Fu} \& {Elkins-Tanton}(2014)}]{2014EPSL.390..128F}
{Fu}, R.~R., \& {Elkins-Tanton}, L.~T. 2014, Earth Planet. Sci. Lett., 390,
  128, \dodoi{10.1016/j.epsl.2013.12.047}

\bibitem[{{Gaillard} {et~al.}(2021){Gaillard}, {Bouhifd}, {F{\"u}ri},
  {Malavergne}, {Marrocchi}, {Noack}, {Ortenzi}, {Roskosz}, \&
  {Vulpius}}]{2021SSRv..217...22G}
{Gaillard}, F., {Bouhifd}, M.~A., {F{\"u}ri}, E., {et~al.} 2021, \ssr, 217, 22,
  \dodoi{10.1007/s11214-021-00802-1}

\bibitem[{{Gerya} \& {Yuen}(2007)}]{2007PEPI..163...83G}
{Gerya}, T.~V., \& {Yuen}, D.~A. 2007, Phys. Earth Planet. Inter., 163, 83,
  \dodoi{10.1016/j.pepi.2007.04.015}

\bibitem[{{Gillmann} {et~al.}(2020){Gillmann}, {Golabek}, {Raymond},
  {Sch{\"o}nb{\"a}chler}, {Tackley}, {Dehant}, \&
  {Debaille}}]{2020NatGe..13..265G}
{Gillmann}, C., {Golabek}, G.~J., {Raymond}, S.~N., {et~al.} 2020, Nat.
  Geosci., 13, 265, \dodoi{10.1038/s41561-020-0561-x}

\bibitem[{{Golabek} \& {Jutzi}(2021)}]{Golabek21}
{Golabek}, G.~J., \& {Jutzi}, M. 2021, \icarus, 363, 114437,
  \dodoi{10.1016/j.icarus.2021.114437}

\bibitem[{{Graham}(2021)}]{Graham21}
{Graham}, R.~J. 2021, arXiv:2104.01224.
\newblock \doarXiv{2104.01224}

\bibitem[{{Grewal} {et~al.}(2021){Grewal}, {Dasgupta}, \&
  {Marty}}]{2021NatAs.tmp...14G}
{Grewal}, D.~S., {Dasgupta}, R., \& {Marty}, B. 2021, Nat. Astron.,
  \dodoi{10.1038/s41550-020-01283-y}

\bibitem[{{Grundy} {et~al.}(2020){Grundy}, {Bird}, {Britt}, {Cook},
  {Cruikshank}, {Howett}, {Krijt}, {Linscott}, {Olkin}, {Parker}, {Protopapa},
  {Ruaud}, {Umurhan}, {Young}, {Dalle Ore}, {Kavelaars}, {Keane}, {Pendleton},
  {Porter}, {Scipioni}, {Spencer}, {Stern}, {Verbiscer}, {Weaver}, {Binzel},
  {Buie}, {Buratti}, {Cheng}, {Earle}, {Elliott}, {Gabasova}, {Gladstone},
  {Hill}, {Horanyi}, {Jennings}, {Lunsford}, {McComas}, {McKinnon}, {McNutt},
  {Moore}, {Parker}, {Quirico}, {Reuter}, {Schenk}, {Schmitt}, {Showalter},
  {Singer}, {Weigle}, \& {Zangari}}]{grundy2020}
{Grundy}, W.~M., {Bird}, M.~K., {Britt}, D.~T., {et~al.} 2020, Science, 367,
  aay3705, \dodoi{10.1126/science.aay3705}

\bibitem[{{Hamano} {et~al.}(2015){Hamano}, {Kawahara}, {Abe}, {Onishi}, \&
  {Hashimoto}}]{2015ApJ...806..216H}
{Hamano}, K., {Kawahara}, H., {Abe}, Y., {Onishi}, M., \& {Hashimoto}, G.~L.
  2015, \apj, 806, 216, \dodoi{10.1088/0004-637X/806/2/216}

\bibitem[{{Harris} {et~al.}(2020){Harris}, {Jarrod Millman}, {van der Walt},
  {Gommers}, {Virtanen}, {Cournapeau}, {Wieser}, {Taylor}, {Berg}, {Smith},
  {Kern}, {Picus}, {Hoyer}, {van Kerkwijk}, {Brett}, {Haldane}, {Fern{\'a}ndez
  del R{\'\i}o}, {Wiebe}, {Peterson}, {G{\'e}rard-Marchant}, {Sheppard},
  {Reddy}, {Weckesser}, {Abbasi}, {Gohlke}, \& {Oliphant}}]{Numpy2020}
{Harris}, C.~R., {Jarrod Millman}, K., {van der Walt}, S.~J., {et~al.} 2020,
  Nature, 585, 357.
\newblock \doarXiv{2006.10256}

\bibitem[{{Helled} {et~al.}(2021){Helled}, {Werner}, {Dorn}, {Guillot},
  {Ikoma}, {Ito}, {Kama}, {Lichtenberg}, {Miguel}, {Shorttle}, {Tackley},
  {Valencia}, \& {Vazan}}]{2021arXiv210308481H}
{Helled}, R., {Werner}, S., {Dorn}, C., {et~al.} 2021, Exp. Astron., in press,
  arXiv:2103.08481.
\newblock \doarXiv{2103.08481}

\bibitem[{Hirschmann(2016)}]{hirschmann2016constraints}
Hirschmann, M.~M. 2016, Am. Mineral., 101, 540, \dodoi{10.2138/am-2016-5452}

\bibitem[{{Hirschmann} {et~al.}(2021){Hirschmann}, {Bergin}, {Blake}, {Ciesla},
  \& {Li}}]{Hirschmann21}
{Hirschmann}, M.~M., {Bergin}, E.~A., {Blake}, G.~A., {Ciesla}, F., \& {Li}, J.
  2021, \pnas, 118, e2026779118, \dodoi{10.1073/pnas.2026779118}

\bibitem[{Hunter(2007)}]{Hunter:2007}
Hunter, J.~D. 2007, Comput. Sci. Eng., 9, 90

\bibitem[{{Johansen} {et~al.}(2014){Johansen}, {Blum}, {Tanaka}, {Ormel},
  {Bizzarro}, \& {Rickman}}]{2014prpl.conf..547J}
{Johansen}, A., {Blum}, J., {Tanaka}, H., {et~al.} 2014, in Protostars and
  Planets VI, ed. H.~{Beuther}, R.~S. {Klessen}, C.~P. {Dullemond}, \&
  T.~{Henning}, 547, \dodoi{10.2458/azu_uapress_9780816531240-ch024}

\bibitem[{{Kane} {et~al.}(2020){Kane}, {Roettenbacher}, {Unterborn}, {Foley},
  \& {Hill}}]{2020PSJ.....1...36K}
{Kane}, S.~R., {Roettenbacher}, R.~M., {Unterborn}, C.~T., {Foley}, B.~J., \&
  {Hill}, M.~L. 2020, PSJ, 1, 36, \dodoi{10.3847/PSJ/abaab5}

\bibitem[{{Kral} {et~al.}(2017){Kral}, {Matr{\`a}}, {Wyatt}, \&
  {Kennedy}}]{kral2017}
{Kral}, Q., {Matr{\`a}}, L., {Wyatt}, M.~C., \& {Kennedy}, G.~M. 2017, \mnras,
  469, 521, \dodoi{10.1093/mnras/stx730}

\bibitem[{{Krijt} {et~al.}(2020){Krijt}, {Bosman}, {Zhang}, {Schwarz},
  {Ciesla}, \& {Bergin}}]{krijt2020}
{Krijt}, S., {Bosman}, A.~D., {Zhang}, K., {et~al.} 2020, \apj, 899, 134,
  \dodoi{10.3847/1538-4357/aba75d}

\bibitem[{{Krijt} {et~al.}(2018){Krijt}, {Schwarz}, {Bergin}, \&
  {Ciesla}}]{krijt2018}
{Krijt}, S., {Schwarz}, K.~R., {Bergin}, E.~A., \& {Ciesla}, F.~J. 2018, \apj,
  864, 78, \dodoi{10.3847/1538-4357/aad69b}

\bibitem[{{Krivov} \& {Wyatt}(2021)}]{krivov2021}
{Krivov}, A.~V., \& {Wyatt}, M.~C. 2021, \mnras, 500, 718,
  \dodoi{10.1093/mnras/staa2385}

\bibitem[{{Li} {et~al.}(2021){Li}, {Bergin}, {Blake}, {Ciesla}, \&
  {Hirschmann}}]{Li+21}
{Li}, J., {Bergin}, E.~A., {Blake}, G.~A., {Ciesla}, F.~J., \& {Hirschmann},
  M.~M. 2021, Sci. Adv., 7, eabd3632, \dodoi{10.1126/sciadv.abd3632}

\bibitem[{{Li} {et~al.}(2019){Li}, {Youdin}, \& {Simon}}]{2019ApJ...885...69L}
{Li}, R., {Youdin}, A.~N., \& {Simon}, J.~B. 2019, \apj, 885, 69,
  \dodoi{10.3847/1538-4357/ab480d}

\bibitem[{{Lichtenberg} {et~al.}(2021{\natexlab{a}}){Lichtenberg}, {Bower},
  {Hammond}, {Boukrouche}, {Sanan}, {Tsai}, \&
  {Pierrehumbert}}]{Lichtenberg2021JGRP}
{Lichtenberg}, T., {Bower}, D.~J., {Hammond}, M., {et~al.} 2021{\natexlab{a}},
  J. Geophys. Res. Planets, 126, e2020JE006711, \dodoi{10.1029/2020JE006711}

\bibitem[{{Lichtenberg} {et~al.}(2021{\natexlab{b}}){Lichtenberg},
  {Dra{\.z}kowska}, {Sch{\"o}nb{\"a}chler}, {Golabek}, \&
  {Hands}}]{2021Sci...371..365L}
{Lichtenberg}, T., {Dra{\.z}kowska}, J., {Sch{\"o}nb{\"a}chler}, M., {Golabek},
  G.~J., \& {Hands}, T.~O. 2021{\natexlab{b}}, Science, 371, 365,
  \dodoi{10.1126/science.abb3091}

\bibitem[{{Lichtenberg} {et~al.}(2019{\natexlab{a}}){Lichtenberg}, {Golabek},
  {Burn}, {Meyer}, {Alibert}, {Gerya}, \& {Mordasini}}]{2019NatAs...3..307L}
{Lichtenberg}, T., {Golabek}, G.~J., {Burn}, R., {et~al.} 2019{\natexlab{a}},
  Nat. Astron., 3, 307, \dodoi{10.1038/s41550-018-0688-5}

\bibitem[{{Lichtenberg} {et~al.}(2019{\natexlab{b}}){Lichtenberg}, {Keller},
  {Katz}, {Golabek}, \& {Gerya}}]{2019EPSL.507..154L}
{Lichtenberg}, T., {Keller}, T., {Katz}, R.~F., {Golabek}, G.~J., \& {Gerya},
  T.~V. 2019{\natexlab{b}}, Earth Planet. Sci. Lett., 507, 154,
  \dodoi{10.1016/j.epsl.2018.11.034}

\bibitem[{{LIFE collaboration} {et~al.}(2021){LIFE collaboration}, {Quanz},
  {Ottiger}, {Fontanet}, {Kammerer}, {Menti}, {Dannert}, {Gheorghe}, {Absil},
  {Airapetian}, {Alei}, {Allart}, {Angerhausen}, {Blumenthal}, {Cabrera},
  {Carri{\'o}n-Gonz{\'a}lez}, {Chauvin}, {Danchi}, {Dandumont}, {Defr{\`e}re},
  {Dorn}, {Ehrenreich}, {Ertel}, {Fridlund}, {Garc{\'\i}a Mu{\~n}oz},
  {Gasc{\'o}n}, {Glauser}, {Grenfell}, {Guidi}, {Hagelberg}, {Helled},
  {Ireland}, {Kopparapu}, {Korth}, {Kraus}, {L{\'e}ger}, {Leedj{\"a}rv},
  {Lichtenberg}, {Lillo-Box}, {Linz}, {Liseau}, {Loicq}, {Mahendra}, {Malbet},
  {Mathew}, {Mennesson}, {Meyer}, {Mishra}, {Molaverdikhani}, {Noack}, {Oza},
  {Pall{\'e}}, {Parviainen}, {Quirrenbach}, {Rauer}, {Ribas}, {Rice},
  {Romagnolo}, {Rugheimer}, {Schwieterman}, {Serabyn}, {Sharma}, {Stassun},
  {Szul{\'a}gyi}, {Wang}, {Wunderlich}, \& {Wyatt}}]{2021arXiv210107500L}
{LIFE collaboration}, {Quanz}, S.~P., {Ottiger}, M., {et~al.} 2021,
  arXiv:2101.07500.
\newblock \doarXiv{2101.07500}

\bibitem[{{Marino} {et~al.}(2016){Marino}, {Matr{\`a}}, {Stark}, {Wyatt},
  {Casassus}, {Kennedy}, {Rodriguez}, {Zuckerman}, {Perez}, {Dent}, {Kuchner},
  {Hughes}, {Schneider}, {Steele}, {Roberge}, {Donaldson}, \&
  {Nesvold}}]{marino2016}
{Marino}, S., {Matr{\`a}}, L., {Stark}, C., {et~al.} 2016, \mnras, 460, 2933,
  \dodoi{10.1093/mnras/stw1216}

\bibitem[{McKinney(2010)}]{pandas:2010}
McKinney, W. 2010, in Proceedings of the 9th Python in Science Conference, ed.
  S.~van~der Walt \& J.~Millman, 51--56

\bibitem[{{McKinnon} {et~al.}(2020){McKinnon}, {Richardson}, {Marohnic},
  {Keane}, {Grundy}, {Hamilton}, {Nesvorn{\'y}}, {Umurhan}, {Lauer}, {Singer},
  {Stern}, {Weaver}, {Spencer}, {Buie}, {Moore}, {Kavelaars}, {Lisse}, {Mao},
  {Parker}, {Porter}, {Showalter}, {Olkin}, {Cruikshank}, {Elliott},
  {Gladstone}, {Parker}, {Verbiscer}, {Young}, \& {New Horizons Science
  Team}}]{mckinnon2020}
{McKinnon}, W.~B., {Richardson}, D.~C., {Marohnic}, J.~C., {et~al.} 2020,
  Science, 367, aay6620, \dodoi{10.1126/science.aay6620}

\bibitem[{Monteux {et~al.}(2018)Monteux, Golabek, Rubie, Tobie, \&
  Young}]{Monteux2018}
Monteux, J., Golabek, G.~J., Rubie, D.~C., Tobie, G., \& Young, E.~D. 2018,
  Space Sci. Rev., 214, 39, \dodoi{10.1007/s11214-018-0473-x}

\bibitem[{{Mousis} {et~al.}(2021){Mousis}, {Aguichine}, {Bouquet}, {Lunine},
  {Danger}, {Mandt}, \& {Luspay-Kuti}}]{mousis2021}
{Mousis}, O., {Aguichine}, A., {Bouquet}, A., {et~al.} 2021, PSJ, 2, 72,
  \dodoi{10.3847/PSJ/abeaa7}

\bibitem[{{Muenow} {et~al.}(1995){Muenow}, {Keil}, \&
  {McCoy}}]{1995Metic..30..639M}
{Muenow}, D.~W., {Keil}, K., \& {McCoy}, T.~J. 1995, Meteoritics, 30, 639,
  \dodoi{10.1111/j.1945-5100.1995.tb01161.x}

\bibitem[{{Mumma} \& {Charnley}(2011)}]{mumma_charnley2011}
{Mumma}, M.~J., \& {Charnley}, S.~B. 2011, \araa, 49, 471,
  \dodoi{10.1146/annurev-astro-081309-130811}

\bibitem[{{Niida} \& {Green}(1999)}]{1999CoMP..135...18N}
{Niida}, K., \& {Green}, D.~H. 1999, Contrib. Mineral. Petrol., 135, 18,
  \dodoi{10.1007/s004100050495}

\bibitem[{{Notsu} {et~al.}(2020){Notsu}, {Eistrup}, {Walsh}, \&
  {Nomura}}]{2020MNRAS.499.2229N}
{Notsu}, S., {Eistrup}, C., {Walsh}, C., \& {Nomura}, H. 2020, \mnras, 499,
  2229, \dodoi{10.1093/mnras/staa2944}

\bibitem[{{{\"O}berg} \& {Bergin}(2021)}]{obergbergin2021}
{{\"O}berg}, K.~I., \& {Bergin}, E.~A. 2021, \physrep, 893, 1,
  \dodoi{10.1016/j.physrep.2020.09.004}

\bibitem[{{{\"O}berg} {et~al.}(2011){{\"O}berg}, {Murray-Clay}, \&
  {Bergin}}]{oberg2011}
{{\"O}berg}, K.~I., {Murray-Clay}, R., \& {Bergin}, E.~A. 2011, \apjl, 743,
  L16, \dodoi{10.1088/2041-8205/743/1/L16}

\bibitem[{{Ormel}(2017)}]{2017ASSL..445..197O}
{Ormel}, C.~W. 2017, {The Emerging Paradigm of Pebble Accretion}, ed.
  M.~{Pessah} \& O.~{Gressel}, Vol. 445, 197,
  \dodoi{10.1007/978-3-319-60609-5_7}

\bibitem[{{Parker}(2020)}]{2020RSOS....701271P}
{Parker}, R.~J. 2020, R. Soc. Open Sci., 7, 201271, \dodoi{10.1098/rsos.201271}

\bibitem[{Piani {et~al.}(2020)Piani, Marrocchi, Rigaudier, Vacher, Thomassin,
  \& Marty}]{Piani+2020}
Piani, L., Marrocchi, Y., Rigaudier, T., {et~al.} 2020, Science, 369, 1110,
  \dodoi{10.1126/science.aba1948}

\bibitem[{{Pinhas} {et~al.}(2019){Pinhas}, {Madhusudhan}, {Gandhi}, \&
  {MacDonald}}]{2019MNRAS.482.1485P}
{Pinhas}, A., {Madhusudhan}, N., {Gandhi}, S., \& {MacDonald}, R. 2019, \mnras,
  482, 1485, \dodoi{10.1093/mnras/sty2544}

\bibitem[{{Price} {et~al.}(2021){Price}, {Cleeves}, {Bodewits}, \&
  {{\"O}berg}}]{price2021}
{Price}, E.~M., {Cleeves}, L.~I., {Bodewits}, D., \& {{\"O}berg}, K.~I. 2021,
  arXiv:2103.12751.
\newblock \doarXiv{2103.12751}

\bibitem[{{Raymond} {et~al.}(2020){Raymond}, {Izidoro}, \&
  {Morbidelli}}]{Raymond2020PlanetaryAstrobiology}
{Raymond}, S.~N., {Izidoro}, A., \& {Morbidelli}, A. 2020, in Planetary
  Astrobiology, ed. V.~S. {Meadows}, G.~N. {Arnay}, B.~E. {Schmidt}, \& D.~J.
  {Des Marais} (University of Arizona Press), 287

\bibitem[{{Reiter}(2020)}]{2020A&A...644L...1R}
{Reiter}, M. 2020, \aap, 644, L1, \dodoi{10.1051/0004-6361/202039334}

\bibitem[{{Rogers} \& {Owen}(2021)}]{2021MNRAS.tmp..530R}
{Rogers}, J.~G., \& {Owen}, J.~E. 2021, \mnras, \dodoi{10.1093/mnras/stab529}

\bibitem[{{Schaefer} \& {Fegley}(2017)}]{2017ApJ...843..120S}
{Schaefer}, L., \& {Fegley}, B.~J. 2017, \apj, 843, 120,
  \dodoi{10.3847/1538-4357/aa784f}

\bibitem[{{Schoonenberg} \& {Ormel}(2017)}]{2017AA...602A..21S}
{Schoonenberg}, D., \& {Ormel}, C.~W. 2017, \aap, 602, A21,
  \dodoi{10.1051/0004-6361/201630013}

\bibitem[{{Singer} {et~al.}(2019){Singer}, {McKinnon}, {Gladman},
  {Greenstreet}, {Bierhaus}, {Stern}, {Parker}, {Robbins}, {Schenk}, {Grundy},
  {Bray}, {Beyer}, {Binzel}, {Weaver}, {Young}, {Spencer}, {Kavelaars},
  {Moore}, {Zangari}, {Olkin}, {Lauer}, {Lisse}, {Ennico}, {New Horizons
  Geology}, Team, {New Horizons Surface Composition Science Theme Team}, \&
  {New Horizons Ralph and LORRI Teams}}]{2019Sci...363..955S}
{Singer}, K.~N., {McKinnon}, W.~B., {Gladman}, B., {et~al.} 2019, Science, 363,
  955, \dodoi{10.1126/science.aap8628}

\bibitem[{{Sossi} {et~al.}(2019){Sossi}, {Klemme}, {O'Neill}, {Berndt}, \&
  {Moynier}}]{2019GeCoA.260..204S}
{Sossi}, P.~A., {Klemme}, S., {O'Neill}, H. S.~C., {Berndt}, J., \& {Moynier},
  F. 2019, \gca, 260, 204, \dodoi{10.1016/j.gca.2019.06.021}

\bibitem[{{Stammler} {et~al.}(2017){Stammler}, {Birnstiel}, {Pani{\'c}},
  {Dullemond}, \& {Dominik}}]{stammler2017}
{Stammler}, S.~M., {Birnstiel}, T., {Pani{\'c}}, O., {Dullemond}, C.~P., \&
  {Dominik}, C. 2017, \aap, 600, A140, \dodoi{10.1051/0004-6361/201629041}

\bibitem[{{Stammler} {et~al.}(2019){Stammler}, {Dra{\.z}kowska}, {Birnstiel},
  {Klahr}, {Dullemond}, \& {Andrews}}]{2019ApJ...884L...5S}
{Stammler}, S.~M., {Dra{\.z}kowska}, J., {Birnstiel}, T., {et~al.} 2019, \apjl,
  884, L5, \dodoi{10.3847/2041-8213/ab4423}

\bibitem[{Tsuchiyama {et~al.}(2021)Tsuchiyama, Miyake, Okuzumi, Kitayama,
  Kawano, Uesugi, Takeuchi, Nakano, \& Zolensky}]{Tsuchi21}
Tsuchiyama, A., Miyake, A., Okuzumi, S., {et~al.} 2021, Sci. Adv., 7, eabg9707,
  \dodoi{10.1126/sciadv.abg9707}

\bibitem[{{Venturini} {et~al.}(2020){Venturini}, {Ronco}, \&
  {Guilera}}]{Venturini20}
{Venturini}, J., {Ronco}, M.~P., \& {Guilera}, O.~M. 2020, \ssr, 216, 86,
  \dodoi{10.1007/s11214-020-00700-y}

\bibitem[{Virtanen {et~al.}(2020)Virtanen, Gommers, Oliphant, Haberland, Reddy,
  Cournapeau, Burovski, Peterson, Weckesser, Bright, {van der Walt}, Brett,
  Wilson, Millman, Mayorov, Nelson, Jones, Kern, Larson, Carey, Polat, Feng,
  Moore, {VanderPlas}, Laxalde, Perktold, Cimrman, Henriksen, Quintero, Harris,
  Archibald, Ribeiro, Pedregosa, {van Mulbregt}, \& {SciPy 1.0
  Contributors}}]{2020SciPy-NMeth}
Virtanen, P., Gommers, R., Oliphant, T.~E., {et~al.} 2020, Nat. Methods, 17,
  261, \dodoi{10.1038/s41592-019-0686-2}

\bibitem[{Waskom {et~al.}(2018)Waskom, Botvinnik, {et~al.}}]{seaborn:2018}
Waskom, M., Botvinnik, O., {et~al.} 2018, Seaborn v0.9.0.
  doi.org/10.5281/zenodo.1313201

\bibitem[{{Wyatt}(2020)}]{wyatt2020}
{Wyatt}, M. 2020, {Extrasolar Kuiper belts}, ed. D.~{Prialnik}, M.~A.
  {Barucci}, \& L.~{Young}, 351--376,
  \dodoi{10.1016/B978-0-12-816490-7.00016-3}

\bibitem[{{Zahnle} {et~al.}(2020){Zahnle}, {Lupu}, {Catling}, \&
  {Wogan}}]{2020PSJ.....1...11Z}
{Zahnle}, K.~J., {Lupu}, R., {Catling}, D.~C., \& {Wogan}, N. 2020, PSJ, 1, 11,
  \dodoi{10.3847/PSJ/ab7e2c}

\bibitem[{{Zeng} {et~al.}(2019){Zeng}, {Jacobsen}, {Sasselov}, {Petaev},
  {Vanderburg}, {Lopez-Morales}, {Perez-Mercader}, {Mattsson}, {Li}, {Heising},
  {Bonomo}, {Damasso}, {Berger}, {Cao}, {Levi}, \&
  {Wordsworth}}]{2019PNAS..116.9723Z}
{Zeng}, L., {Jacobsen}, S.~B., {Sasselov}, D.~D., {et~al.} 2019, Proc. Nat.
  Acad. Sci. USA, 116, 9723, \dodoi{10.1073/pnas.1812905116}

\bibitem[{{Zhang} {et~al.}(2019){Zhang}, {Bergin}, {Schwarz}, {Krijt}, \&
  {Ciesla}}]{zhang2019}
{Zhang}, K., {Bergin}, E.~A., {Schwarz}, K., {Krijt}, S., \& {Ciesla}, F. 2019,
  \apj, 883, 98, \dodoi{10.3847/1538-4357/ab38b9}

\end{thebibliography}
\bibliographystyle{aasjournal}

\end{document}